\documentclass[a4paper, 12pt]{article}
\usepackage{epsfig}
\usepackage{epic}
\usepackage{color}
\usepackage{graphicx}
\usepackage{algorithm}
\usepackage{algorithmic}
\usepackage{setspace}
\usepackage{amssymb,amsmath,amsthm}

\begin{document}

\title{ Chains of Mean Field Models}
\author{S.Hamed Hassani, Nicolas Macris and Ruediger Urbanke
\\
\\
\small{Laboratory for Communication Theory}\\
\small{School of Computer and Communication Science}\\
\small{Ecole Polytechnique F\'ed\'erale de Lausanne}\\
\small{Station 14, EPFL, CH-1015 Lausanne, Switzerland}}
\date{}
\maketitle

\begin{abstract} 
\noindent 
We consider a collection of Curie-Weiss (CW) spin systems, possibly with a random field, each of which is placed 
along the positions of a one-dimensional chain. The CW systems are   
coupled together by a Kac-type interaction in the longitudinal direction of the chain and by an infinite range interaction in the direction transverse 
to the chain. Our motivations for studying this model come from recent findings in the theory of error correcting codes 
based on spatially coupled graphs. We find that, although much simpler than the codes, the model studied here 
already displays similar behaviors.
We are interested in the van der Waals curve in a 
regime where the size of each Curie-Weiss model tends to infinity, and the length of the chain  
and range of the Kac interaction are large but finite. 
Below the critical temperature, and with appropriate boundary conditions, there appears a series 
of equilibrium states representing kink-like interfaces between the two equilibrium states
of the individual system.
The van der Waals curve oscillates periodically
around the Maxwell plateau. These oscillations have a period inversely proportional to the chain length 
and an amplitude exponentially small in the range of 
the interaction; in other words the spinodal points of the chain model lie exponentially close to the
phase transition threshold. The amplitude of the oscillations is closely related to a Peierls-Nabarro 
free energy barrier for the motion of the kink along the chain.
Analogies to similar phenomena and their possible algorithmic significance for graphical models of 
interest in coding theory and theoretical computer science are pointed out. 
\end{abstract}

\section{Introduction}\label{section 1}


Low-Density Parity-Check (LDPC) codes \cite{Ruediger-Tom-Book} are a class of parity check codes designed from appropriate ensembles of sparse random 
graphs. These have emerged as fundamental building blocks of modern error correcting schemes for communication over noisy channels.
Their great advantage is the existence of efficient, low complexity, decoding algorithms. It is quite remarkable that these systems
can be viewed as mean field spin glasses on random 
sparse graphs. This connection has been known for some years and it is recognized that it is quite far reaching. 
The noise thresholds for the transition between reliable and unreliable communication 
are obtained by a performance 
curve\footnote{called Extended Belief Propagation Generalized Extrinsic Information Transfer curve} 
analogous to a van der Waals isotherm. Analogs of spinodal points 
determine thresholds - called Belief Propagation thresholds - under the efficient low complexity Belief Propagation decoding algorithm. 
The first order phase transition threshold - called Maximum a Posteriori threshold - can be obtained by a Maxwell construction and determines 
the threshold associated to optimal but computationally impractical decoding. We refer to
\cite{Ruediger-Tom-Book} and \cite{Mezard-Montanari} for further information and background on both the coding theory and statistical mechanics aspects.

In order to design good codes one may try to design sparse graph ensembles such that the spinodal 
points are separated from the Maxwell plateau by a small gap. It has been realized recently \cite{Lentmaier-Fettweis-Zigangirov-Costello},
\cite{Kudekar-Richardson-Urbanke-II}, \cite{Kudekar-Richardson-Urbanke-I}  that this goal can be achieved - in a versatile way - by a class of 
so-called {\it convolutional terminated LDPC codes}, which were first introduced in the early works \cite{Felstrom-Zigangirov},
\cite{Engdahl-Zigangirov},
\cite{Lentmaier-Truhachev-Zigangirov}. They can be viewed as {\it chains of individual 
LDPC codes} of length $n$, that are coupled, along a one-dimensional spatial direction of length $2L+1$, across a coupling window of size $w$ covering many individual codes. 
In the regime where $n>>L>>w>>1$ and with appropriate boundary conditions at the ends of the chain, the performance of the Belief Propagation decoder is excellent.
In particular the Belief Propagation threshold improves and saturates towards the Maximum a Posteriori threshold, which is the best possible value. 
In statistical mechanical parlance the spinodal points 
come infinitely close to the Maxwell plateau. It has also been observed that the performance curve of the spatially coupled code ensemble displays a fine 
oscillating structure, with oscillations of period $O(1/L)$ as the Maxwell plateau is approached. In the coding theoretic context these observations go under the name of 
{\it threshold saturation phenomenon}.

In order to better understand the fundamental origins of threshold saturation we have investigated a wide variety of models that are 
chains of spatially coupled mean field systems, with appropriate boundary conditions. These include chains of Curie-Weiss models, random constraint satisfaction problems 
such as $K$-SAT, $Q$-coloring and we have found that it is a very general phenomenon. See \cite{Hassani-Macris-Urbanke-I} for a short summary and 
the conclusion for further discussions of these aspects. 
It has also been shown to occur in various 
other settings such as muti-user communication and compressed sensing (see for example \cite{tanaka-kawabata}, \cite{kudekar-kasai}, \cite{pfister-kudekar},
\cite{mezardetal}).

In the present work we 
present in detail what we believe is the simplest and clearest situation that captures 
the basic underpinnings of threshold saturation. We introduce a one dimensional chain of $2L+1$ Curie-Weiss\footnote{Ising model on a complete graph} (CW) 
spin systems coupled together by an interaction which is local in the longitudinal (or chain) direction and infinite range
in the transverse direction. The local interaction 
is of Kac type with an increasing range and inversely decreasing intensity, and is ferromagnetic. 
This model can be viewed as an anisotropic 
Ising system with a Kac interaction along one longitudinal direction and a Curie-Weiss infinite range interaction along the ''infinite dimensional`` transverse direction. 
We also analyze a variant of this model, where the individual system is a Random Field Curie-Weiss (RFCW) model.                                                                                                                                                       

\begin{figure}[h!]
\begin{centering}
\input{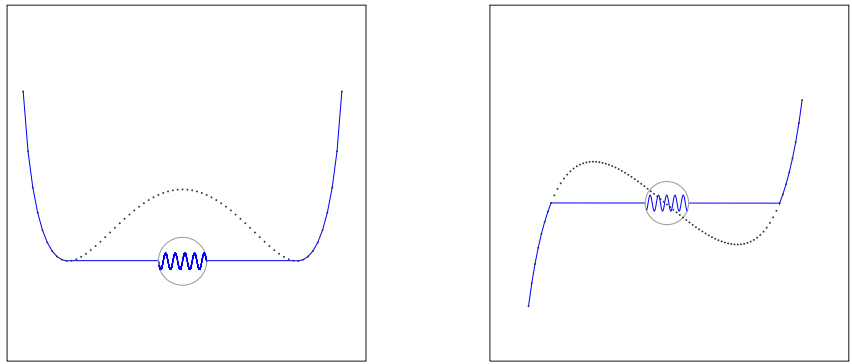}
\caption{\small{Qualitative illustration of main result. Dotted curves: free energy and van der Waals isotherm of the single system for a coupling strength $J>1$ ($J=1$ is the critical point).
Continuous curves: free energy and  
van der Waals isotherm of the coupled chain for $2L>>w>>1$. 
The oscillations extend 
throughout the plateau with a period $M/2L$ and amplitudes $O(L^{-1}e^{\frac{2\alpha\pi^2 w}{JM}})$ (left) and 
$O(e^{\frac{2\alpha\pi^2 w}{JM}})$ (right) where $M=\textrm{width of plateau}$, $\alpha=O(1)$ depends on the details of the interaction
(sect. \ref{section-4}).
Close to the end points of the plateau, within a distance $O(L^{-1/2})$, boundary effects are important and the curves 
depend on details of the boundary conditions (sect. \ref{section-3}).}
}
\label{fr-vdw}
\end{centering}
\end{figure}

The main focus of this 
paper is to understand the evolution of the van der Waals isotherm of the coupled chain when the individual underlying system is infinite and, 
the range $w$ of the Kac 
interaction and the longitudinal length $2L+1$ 
become large $L>>w>>1$ but are still finite. 
This problem is studied for temperatures below the critical temperature of the individual system. The magnetizations at the boundaries are set equal 
to the two equilibrium states of the individual system\footnote{This is the right analogy with the coding theory context. Other choices are not relevant to the 
threshold saturation phenomenon.}. 
In the limit where both $L$ and $w$
become infinite the {\it van der Waals} isotherm of the coupled chain tends to the {\it Maxwell} isotherm of the individual CW 
system. In particular the spinodal points of the coupled chain approach the Maxwell plateau of the individual system: this is the threshold saturation phenomenon.
Correspondingly the canonical free energy of the coupled chain is given by the convex envelope of the individual CW model. 
When $L$ and $w$ are large but remain finite, below the critical point of the CW model,
a fine structure develops around the Maxwell plateau: the straight line is 
replaced by an oscillatory curve with period of the order of the inverse chain length and amplitude exponentially small in the range 
of the Kac interaction (see fig. \ref{fr-vdw} and formulas \eqref{main-result-2}, \eqref{main-result-3}). Correspondingly, the finite-size corrections 
to the canonical free energy display, in addition to a ``surface tension'' shift, the 
same oscillations
along the line joining the two equilibrium states of 
the individual system (see fig. \ref{fr-vdw} and formula \eqref{main-result-1}). The series of stable minima is in  
correspondence with kink-like 
magnetization density profiles, representing the coexistence of the two stable phases of the individual system, 
with a well localized interface centered 
at successive positions of the chain (formulas \eqref{kinky}, \eqref{specific}, and fig. \ref{const-J=1.1}). 
A series of unstable maxima is associated
 with kinks centered in-between 
successive positions. 
The amplitude of the oscillations can be interpreted as a Peierls-Nabarro free energy barrier for the motion of a kink along the chain.
We point out that although our analytical results are for the regime of large $w$, numerically we very clearly observe the same phenomena even when $w=1$ which 
corresponds to nearest neighbored coupling between individual CW systems (section \ref{section-3}).

One of the virtues of the present simple model is that it can, to a large extent, be treated analytically by rather explicit methods. While our analysis is not entirely 
rigorous, we believe that it can be made so. We have refrained to do so here, in order that the mathematical technicalities do not obscure 
the main picture.

In view of the classical work of Lebowitz and Penrose \cite{Lebowitz-Penrose} it is perhaps not surprising
that the van der Waals curve of the chain tends to the Maxwell isotherm. However there  is 
a difference: in \cite{Lebowitz-Penrose}
 their ``reference system'' has 
a short range potential whereas here the individual CW system has infinite range interaction. 
Therefore in \cite{Lebowitz-Penrose} the sequence of infinite volume canonical free energies {\it remains convex} during the Kac limiting process.
This is not the case in our setting (below the critical point).
Similarly, the present limit is not equivalent to anisotropic 
Kac limits. Again, during such limiting processes the sequence of infinite volume free energies remains convex 
whereas in the present model the 
convergence proceeds through oscillatory curves of ever smaller amplitude. 
We are not aware if anisotropic Kac limits have been discussed in the literature and we make these remarks more precise
in section \ref{section-6}.

There is a large literature on models involving a mixture of infinite range (mean field) and
short range interactions; we point to \cite{perk}, \cite{susuki} for the interested reader.  Here we wish to point out a few works 
that are more specifically related to the present model. Falk and Ruijgrok introduced a 
chain of spin systems
where every spin interacts only with all spins of the neighboring chains \cite{falk-ruijgrok}.
The model and mean field equations of this model appear to be similar to ours \cite{falk-ruijgrok}, \cite{thompson}; but one crucial difference 
comes from the fact that there are no intra-chain interactions in their model, and thus the individual system has a trivial isotherm\footnote{The model was 
introduced with 
different motivations in mind. Namely to establish how the critical temperature ``deteriorates`` as the length of the chain grows.}. 
A mean field {\it approximation}, related to our model, has been used to analyze the phase diagram of the  
two and three dimensional Axial Next Nearest Neighbor Interaction (ANNNI) Ising model
\cite{Bak-Boehm}, \cite{Bak}. The motivation there was completely different than ours and was focused on 
the existence of incommensurate phases. In \cite{krug-lebowitz-spohn-zhang} stationary nonequilibrium states of the van Beijeren-Schulman 
model \cite{vBS} of a stochastic lattice gas are studied. These are controlled by a ''free energy functional`` that bears structural similarities 
with the present equilibrium model for $w=1$. 
Another relation is with the 
works of \cite{Cassandro}, \cite{Masi-Orlandi-Presutti-Triolo-I}, \cite{Penrose}
on the coexistence of two phases in a strictly one dimensional Ising model with Kac interaction. 
Static as well as dynamical aspects have been 
studied and in appropriate hydrodynamic limits the magnetization density satisfies a mean 
field equation which is (at least for Glauber dynamics) a 
continuous version of the discrete mean field equation derived here for the chain model.

In section \ref{section-2} we set up our basic model and give a formal solution. 
The asymptotic analysis for $L>>w>>1$ is performed in section \ref{section-4} and this is supplemented by numerical 
simulations valid for all $w$ in section \ref{section-3}. Section \ref{section-5} contains a generalization to
a model with random fields. Section \ref{section-6} discusses the differences with anisotropic Kac limits and in \ref{section-7} 
we point out further analogies with 
error correcting codes and models of constraint satisfaction problems.

\section{Chain of Ising systems on complete graphs}\label{section-2}

\subsection{Curie-Weiss model}\label{section2.1}

We start with a brief review of standard material about
 the Curie-Weiss 
model (CW) in the canonical ensemble (or lattice-gas interpretation) which is the natural setting for our purpose. The Hamiltonian is
\begin{equation}\label{hamiltonian}
H_N = -\frac{J}{N}\sum_{\langle i,j\rangle} 
s_i s_j,
\end{equation}
where the spins $s_i=\pm 1$ are attached to the $N$ vertices 
of a complete graph. In \eqref{hamiltonian} the sum over $\langle i,j\rangle$ carries 
over all edges of the graph and 
we take a ferromagnetic 
coupling $J>0$. In the sequel we absorb the 
inverse temperature in this parameter. 
The free 
energy, for a fixed magnetization $m=\frac{1}{N}\sum_{i=1}^N s_i$, is
\begin{equation}
\Phi_N(m) = -\frac{1}{N}\ln Z_N, 
\qquad  Z_N = \sum_{s_i: m=\frac{1}{N}\sum_{i=1}^N s_i} e^{- H_N}
\end{equation}
It has a well defined thermodynamic 
limit (we drop an irrelevant additive constant)
\begin{equation}\label{varia} 
\lim_{N\to+\infty}\Phi_N(m)\equiv\Phi(m)=-\frac{ J}{2}m^2 - \mathcal{H}(m) 
\end{equation}
equal to the internal energy $-\frac{ J}{2}m^2$ minus the binary entropy,  
\begin{equation}
\mathcal{H}(m) = -\frac{1+m}{2}\ln \frac{1+m}{2} - 
\frac{1-m}{2}\ln \frac{1-m}{2},
\end{equation}
of configurations with total magnetization 
$m$. In the canonical formalism the equation of state is simply
\begin{equation}\label{single-VdW}
h = \frac{\partial \Phi(m)}{\partial m}=-Jm + \frac{1}{2}\ln\frac{1+m}{1-m},
\end{equation}
which is equivalent to the Curie-Weiss mean field equation
\begin{equation}\label{cuwe}
m = \tanh(Jm + h).
\end{equation}

As is well known,
from the van der Waals curve $h(m)$ \eqref{single-VdW}, one can derive an equation of state that satisfies thermodynamic stability
requirements  
from a Maxwell construction. Similarly a physical free energy is given by the convex envelope of \eqref{varia}.
\begin{figure}[htp]\label{fig:Van_der_waal}
\begin{centering}
\input{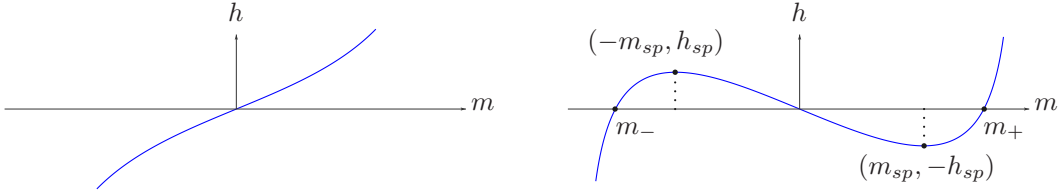}
\caption{\small{Left: van der Waals curve in the high temperature phase $J<1$. Right: low temperature phase $J>1$.
For $m\notin (m_{-}, m_{+})$ the curve describes stable equilibrium states and 
for $m\in (m_{-}, -m_{sp})\cup (m_{sp}, m_{+})$ metastable states.
For $m\in (-m_{sp}, m_{sp})$ the system is unstable. The Maxwell plateau describes superpositions of $m_{-}$ and $m_{+}$ states.}
}
\end{centering}
\end{figure}
For $J\leq 1$, $h(m)$ is monotone 
(see fig. \ref{fig:Van_der_waal}) and the inverse relation
$m(h)$ yields the thermodynamic equilibrium magnetization at a given external magnetic field. For $J>1$ \eqref{single-VdW}-\eqref{cuwe} 
may have more than one solution for a given $h$ (see fig. \ref{fig:Van_der_waal}). 
Starting with $h$ positive and large, we follow a branch $m_{+}(h)$ corresponding to a thermodynamic 
equilibrium state
till the point $(h=0_+, m=m_{+})$. Then we follow a lobe corresponding to 
a metastable state 
till the spinodal point $(h=-h_{sp}, m=m_{sp})$ at the minimum of 
the lobe. Finally from the spinodal point to the origin the curve corresponds to an unstable state 
(where $\frac{\partial^2\Phi(m)}{\partial m^2}<0$).
The situation is symmetric if we start on the other side of 
the curve with $h$ 
large negative. We first follow a stable equilibrium state with magnetization equal to 
$m_{-}(h)$  till the point $(h=0_{-}, m=m_{-})$; we then follow a metastable state till
the left spinodal point $(h=h_{sp}, m=-m_{sp})$; and finally an unstable state till the origin. 

The following expressions valid for $J>1$, will be useful in the sequel,
\begin{equation}\label{spin}
\begin{cases}
 h_{sp} &= -\sqrt{J(J-1)} +\frac{1}{2}\ln\frac{J+\sqrt{J-1}}{J-\sqrt{J-1}}
\approx \frac{1}{3} (J-1)^{\frac{3}{2}}, 
\\
m_{sp} &= \sqrt\frac{J-1}{J}\approx \sqrt{J-1},
\end{cases}
\end{equation}
and 
\begin{equation}
m_{\pm} \approx \pm\sqrt{3(J-1)}.
\end{equation}
In these formulas $\approx$ means that $J\to 1_{+}$. 
The first order phase transition line is $(h_c=0, J>1)$ and terminates at the critical 
second order phase transition point
$(h_c=0, J=1_+)$. For $J<1$ and $h=0$, $m_{\pm}=0$. We will see that for the chain models the difference 
between the first order phase transition and spinodal thresholds becomes much smaller,
and in fact vanishes exponentially fast with the width of the coupling along the chain.

\subsection{Chain Curie-Weiss model}\label{section2.2}

Consider $2L+1$ integer positions $z=-L,...,+L$ on a one dimensional line. 
At each position we attach a single CW
spin system. The spins of each system are labeled as $s_{iz}$, $i=1,...,N$, and 
are subjected to 
a magnetic field $h$. The spin-spin coupling is given by 
\begin{equation}\label{spin-spin-coupling}
-\frac{1}{N} J_{z,z^\prime} s_{iz} s_{jz^\prime} =-\frac{J}{Nw}
g(w^{-1}\vert z-z^\prime\vert)s_{iz}s_{jz^\prime}
\end{equation}
where the function $g(\vert x\vert )$ satisfies the following requirements:
\vskip 0.25cm
\indent a)  
It takes non-negative values and is independent of $i,j$ and $L$. It may 
depend on $w$ itself (see comments below) however we still write $g(\vert x\vert)$ instead of $g_{w}(\vert x\vert)$.
\vskip 0.25cm
\indent b)  Has finite support $[-1,+1]$, i.e $g(\vert x\vert)=0$ for $\vert x\vert > 1$.
\vskip 0.25cm
\indent c) It satisfies the normalization condition
\begin{equation}
 \frac{1}{w}\sum_{z=-\infty}^{+\infty} g(w^{-1}\vert z\vert) = 1.
\end{equation}
This is a purely ferromagnetic interaction which is of 
Kac type in the one dimensional $z$ direction and is purely mean field in the transverse 
''infinite dimensional`` direction. Condition a) ensures that we can find
  asymptotically (as $z\to \pm \infty$)
translation invariant states. Allowing for sign variations certainly leads to a richer phase diagram 
and is beyond the scope of this paper. Conditions b) and c) can easily
be weakened without changing the main results
at the expense of a slightly more technical analysis. One could allow for functions that have infinite support
and decay fast enough (with finite second moment) at infinity. 
The normalization condition is set up so that the strength of the total 
coupling of one spin to the rest 
of the system equals $J$ as $N\to +\infty$ (as in the individual CW system).
For any given function $\tilde g(\vert x\vert)$ that is summable, 
we can always construct one that satisfies this condition 
$g(\vert x\vert )= w\tilde g(\vert x\vert )/ \sum_{z=-\infty}^{+\infty} 
\tilde g(w^{-1}\vert z\vert)$. This means that
 in general $g(\vert x\vert)$ will depend explicitly on $w$; however we could relax this 
slight fine tuning by taking 
the normalization condition to hold only asymptotically as $w\to+\infty$, namely that 
$\int_{-\infty}^{+\infty} g(\vert x\vert ) = 1$.

The Hamiltonian is 
\begin{equation}\label{chain-ham}
H_{N,L}  = 
-\frac{1}{N}\sum_{\langle iz,jz^\prime\rangle} J_{z,z^\prime}
s_{iz} s_{jz^\prime}.
\end{equation}
The first sum carries over all pairs $\langle iz,jz^\prime\rangle$ (counted once each) 
with $i,j=1,...,N$ and $z,z^\prime=-L,...,L$. We will adopt a canonical ensemble with 
\begin{equation}\label{canoni}
 m= \frac{1}{(2L+1)N}\sum_{i=1, z=-L}^{N, L} s_{iz}
\end{equation}
fixed.
The partition function $Z_{N,L}$ is defined by summing $e^{-H_{N,L}}$ over 
all spin configurations 
$\{s_{iz}=\pm 1, i=1,...,N; z=-L,...,L\}$ satisfying \eqref{canoni}. 

We now show that the free 
energy
$f_{N,L}=-\frac{1}{N(2L+1)}\ln Z_{N,L}$ 
is given by a variational principle. 
Let us introduce a magnetization density at position $z$
\begin{equation}
m_z = \frac{1}{N}\sum_{i=1}^N s_{iz},
\end{equation}
and a matrix
\begin{equation}\label{laplacian}
D_{z,z^\prime} =  J_{z,z^\prime} - J\delta_{z,z^\prime}.   
\end{equation}
This matrix is symmetric and for any $z^\prime=-L,...,+L$ it satisfies 
\begin{equation}\label{laplacian-sum}
\sum_{z=-L}^{L} D_{z,z^\prime}
\leq J\,
\mathbb{I}(\vert z^\prime \pm L\vert \leq w)
\end{equation}
The important point here is that the row sum of \eqref{laplacian} 
vanishes except for $z^\prime$ close to the boundaries.
In this respect  one may think of \eqref{laplacian} as 
one-dimensional Laplacian matrix and, as we will see, this 
becomes exactly the case in an appropriate continuum limit of the model.
The Hamiltonian can be re-expressed as (up to a constant)
\begin{equation}\label{hamham}
 H_{N,L}=-\frac{N}{2}\sum_{z,z^\prime=-L}^L D_{z,z^\prime}
m_{z} m_{z^\prime} - \frac{NJ}{2}\sum_{z=-L}^L m_z^2
\end{equation}
In the thermodynamic limit the magnetization density
becomes a continuous variable $m_z\in[-1,+1]$ and the partition 
sum becomes (up to irrelevant prefactors)
\begin{align}
Z_{N,L} = 
&
\int_{[-1,+1]^{2L+1}} \prod_{z=-L}^{L} dm_z
\,\,
\delta\biggl((2L+1)m - \sum_{z=-L}^L m_z\biggr) 
\nonumber
\\
&
\,\times
\exp{-N\biggl(-\frac{1}{2}\sum_{z,z^\prime=-L}^L D_{z,z^\prime}
m_{z} m_{z^\prime}
+\Phi(m_z)\biggr)}.
\label{compact-spins}
\end{align}
This integral can be interpreted as the canonical partition function of a
one dimensional chain of 
continuous compact spins $m_z\in [-1,+1]$, at nearly zero temperature $N^{-1}$,  
with Hamiltonian  
\begin{equation}
\Phi_{L}[\{m_z\}] = -\frac{1}{2}\sum_{z,z^\prime=-L}^L D_{z,z^\prime} m_z m_{z^\prime} 
+ \sum_{z=-L}^L \Phi(m_z).
\end{equation}
The free energy of the 
finite chain obtained from \eqref{compact-spins} is
\begin{align}\label{cano-free-energy}
 F_L(m)&=-\lim_{N\to+\infty}\frac{1}{N}\ln Z_{N,L}
= \min_{m_z:\sum_z m_z = (2L+1)m}\Phi_{L}[\{m_z\}].
\end{align}

The solutions of this variational problem satisfy the set of equations
\begin{equation}\label{var}
\begin{cases}
\sum_{z^\prime=-L}^{L} D_{z,z^\prime} m_{z^\prime}  = 
\Phi^\prime(m_z) -\lambda
\\
\,\,\, m  = \frac{1}{2L+1}\sum_{z=-L}^{L} m_z,
\end{cases} 
\end{equation}
were $\lambda$ is a Lagrange multiplier associated to the constraint 
(and where $\Phi^\prime$ denotes the derivative of the function $\Phi$). 
Denote by $(\lambda^*, m_z^*)$ a solution of \eqref{var} for given $m$. The van der Waals  
equation of state is then given by the usual thermodynamic relation
\begin{equation}\label{vandercan}
h=\frac{1}{2L+1}\frac{\partial F_{L}(m)}{\partial m}.
\end{equation}
In fact $h=\lambda^*$. Indeed,
differentiating in \eqref{vandercan} thanks to the chain rule 
and then using \eqref{var} yields,
\begin{align}\label{one}
h
&
=\frac{1}{2L+1}\sum_{z=-L}^L\biggl(-\sum_{z^\prime=-L}^L D_{z,z^\prime} m_{z^\prime}^* 
+ \Phi^\prime(m_z^*)\biggr) \frac{d m_z^*}{d m}
\nonumber \\ &
=
\frac{\lambda^*}{2L+1}\sum_{z=-L}^L \frac{d m_z^*}{d m}
\nonumber \\ &
=\lambda^* 
\end{align}
Let us make a few remarks on alternative forms for the above equations.
First, summing over $z$ the first equation in \eqref{var} we obtain thanks to \eqref{laplacian-sum}
\begin{equation}\label{two}
 h = \frac{1}{2L+1}\sum_{z=-L}^{L} \Phi^\prime(m_z^*) + O(\frac{w}{L})
\end{equation}
Second, using the explicit expression for the potential $\Phi(m_z)$, equation \eqref{var}
for the minimizing profiles can be cast in the form
\begin{equation}\label{crit}
\begin{cases}
 m_z^*= \tanh\bigl\{Jm_z^* +h +\sum_{z^\prime=-L}^{+L} D_{z,z^\prime}  
m_{z^\prime}^*\bigr\},\\
m  = \frac{1}{2L+1}\sum_{z=-L}^{L} m_z^*.
\end{cases}
\end{equation}
This is a generalization of the CW equation to the chain model. 
We discuss a continuum version of the equation in the next section.
We also remark that this is the form of the equation which has a convenient 
generalization for the chain with random fields (see section \ref{section-5}). 

For $J\leq 1$ the single CW system has a unique equilibrium magnetization so we expect 
a unique translation invariant solution 
for \eqref{crit}, namely $m_z^* = m$ (neglecting boundary effect). It then follows that the 
van der Waals curve of the chain model is the same as that of 
the single CW model. On the other hand for $J>1$ the solutions of \eqref{var} display
non-trivial kink-like magnetization profiles. These solutions are 
responsible for an interesting
oscillating structure in the van der Waals curve.
This is investigated both numerically and to some extent analytically
in the next two sections.

Before closing this section we want to point out that the same system can be analyzed 
in the grand-canonical ensemble (always from the lattice gas perspective) by adding
an external magnetic field term $-h\sum_{i,z} s_{iz}$ to the Hamiltonian \eqref{hamham}.
The definition of
the model is completed by imposing the boundary conditions:
\begin{equation}\label{bcd}
\frac{1}{N}\sum_{i=1}^N s_{i,\pm L} = m_{\pm}(h), 
\end{equation}
where $m_{\pm}(h)$ are the local minima of $\Phi(m)-h m$. 
Note that when the minimum
is unique (for $J\leq 1$ or $J>1$ and $\vert h\vert\geq h_{sp}$) 
the two boundary conditions $m_{\pm}(h)$ are simply equal. The free energy 
(or minus the pressure of the lattice gas) is given by the variational problem
\begin{equation}\label{grand-canon-free}
\min_{m_z: m_{\pm L} = m_{\pm}(h)} 
\biggl(-\frac{1}{2}\sum_{z,z^\prime=-L}^L D_{z,z^\prime} m_z m_{z^\prime} 
+ \sum_{z=-L}^{L} (\Phi(m_z)-h m_z)\biggr)
\end{equation}
The critical points of this functional satisfy
\begin{equation}\label{critical}
\begin{cases}
\sum_{z^\prime=-L}^{+L} D_{z,z^\prime}  m_{z^\prime}  = \Phi^\prime(m_z) - h
\\
\, m_{\pm L}  = m_{\pm}(h)
\end{cases}
\end{equation}
which is also equivalent to
\begin{equation}\label{cri}
\begin{cases}
 m_z= \tanh\bigl\{Jm_z +h +\sum_{z^\prime=-L}^{+L} D_{z,z^\prime}  m_{z^\prime}\bigr\}\\
\, m_{\pm L}  = m_{\pm}(h).
\end{cases}
\end{equation}
The 
 solutions 
of \eqref{critical} or \eqref{cri} define curves $m_z^*(h)$. 
Proving the existence of these curves is beyond our scope here; in general these
 are not single 
valued because the solutions are not unique for a given $h$.
The van der Waals relation $h(m)$ can be recovered from these curves by using
\begin{equation}\label{total-mag}
 m = \frac{1}{2L+1}\sum_{z=-L}^{L} m_z^*(h)
\end{equation}

The magnetization profiles of the canonical and grand-canonical ensembles only differ near the boundaries. Their bulk behavior which 
is our interest are identical. In this paper this is verified numerically (section \ref{section-3}).
In the next section we find it more convenient to refer to the grand-canonical formalism \eqref{critical}, \eqref{cri}, \eqref{total-mag}.

\section{A continuum approximation}\label{section-4}

The asymptotic limit of $L>>w>>1$ reduces the 
solution of equations \eqref{critical}, \eqref{cri}, \eqref{total-mag} to a problem of 
Newtonian mechanics.   
In this limit we obtain a non-linear integral equation 
 which cannot be solved exactly; but 
whose solutions can be qualitatively
discussed for any fixed $J>1$ (an exact solution for all $J>1$ is provided in a special case). 
Near the critical point $J\to 1_+$ this equation is solved and the
solutions used to compute an approximate version of the van der Waals curve. In this way all the features
of the numerical solution are reproduced.
Usually continuum limits are obtained when
a lattice spacing $a$ between neighboring sites of the chain is sent to zero. This set up can also be explored for 
the present model and one finds that it is non trivial only near the critical point $J\to 1_+$, where it yields qualitatively
identical results to the limit $w\to+\infty$, $J\to 1_+$.  
Away from the critical point ($J>1$) $a\to 0$ is a trivial limit which 
supports only homogeneous states, contrary to the  
$w\to +\infty$ limit which displays non trivial features for all $J>1$.
\vskip 0.25cm
\noindent{\bf Asymptotics for $L>>w>>1$.}
We set
\begin{equation}
 z=w x,\qquad m_z = m_{w x}\equiv\mu(x)
\end{equation}
so 
equation \eqref{critical} is equivalent to
\begin{equation}\label{left}
\frac{J}{w}\sum_{z^\prime=-L}^L \biggl\{g(\vert x-\frac{z^\prime}{w}\vert)
- w\delta_{x, \frac{z^\prime}{w}}\biggr\}\mu\biggl(\frac{z^\prime}{w}\biggr) 
= \Phi^\prime(\mu(x))-h.
\end{equation}
We take the limits $L\to+\infty$ first and $w\to +\infty$ second, so that this equation becomes
\begin{equation}\label{becomes}
 J\int_{-\infty}^{+\infty} dx^\prime \bigl\{g(\vert x^\prime\vert) - 
\delta(x^\prime)\bigr\}\mu(x+x^\prime) = \Phi^\prime(\mu(x))-h.
\end{equation}
which can also be cast in a more elegant form ($*$ denotes convolution)
\begin{equation}\label{form}
 \tanh (J g*\mu +h) = \mu.
\end{equation}
We cannot solve this equation in general, except for the special case of uniform $g$. Equ. \eqref{form} for $h=0$
appears in \cite{Cassandro}, \cite{Masi-Orlandi-Presutti-Triolo-I} and existence plus properties of solutions has been discussed.
For our purpose a qualitative discussion of its solutions suffices and we briefly outline it for the reader's convenience.
For $\vert x\vert >> 1$ we can expand
$\mu(x+x^\prime)$ to second order (in \eqref{becomes}) since $g(\vert x\vert)$ vanishes 
for $\vert x\vert >1$. This yields the approximate equation 
\begin{equation}
 J\kappa\mu^{\prime\prime}(x) \approx \Phi^\prime(\mu(x))-h,
\qquad \kappa = \frac{1}{2}\int_{-\infty}^{+\infty}dx^\prime\, 
 x^{\prime 2} g(\vert x^\prime\vert).
\end{equation}
We recognize here Newton's second law for a particle moving 
in the {\it inverted potential} $-\Phi(\mu(x))$ where $\mu(x)$ is the particle's position 
at time $x$ and $J\kappa$ its mass. 
Note this is not a Cauchy problem 
with fixed initial position and velocity, 
but a boundary value problem with $\lim_{x\to\pm\infty}\mu(x) = m_{\pm}(h)$; 
the boundary conditions automatically fix the initial and final velocities.  
The nature of the solutions can be deduced by applying the conservation of mechanical
energy for a ball rolling in the inverted potential. 
For $J<1$ the inverted potential has a single maximum at $m_+(h)=m_-(h)$ and the only 
solution is
$\mu(x)=m_{\pm}(h)$, corresponding to a homogeneous state. 
In fact this is also true for the integral equation.  Now we consider $J>1$ and $h=0$.
At time $-\infty$ the particle is 
on the left maximum and starts rolling down infinitely slowly, then spends a finite 
time in the bottom of the potential well, and finally 
climbs to the right maximum  infinitely slowly to reach it at time $+\infty$. 
For the magnetization
profile $m_z$ this translates to a kink-like state. Note that the center of the kink is set by the normalization 
condition \eqref{total-mag}, and thus we have a continuum of solutions parametrized by the parameter
 $m$ on the Maxwell plateau $[m_-, m_+]$. 
For $J>1$ and $h> 0$, the particle starts 
with a positive initial velocity,
rolls down the potential well, and finally reaches the right maximum infinitely slowly. Thus 
$\mu(x)=m_+(h)$ for all $x$ except for an interval of width $O(1)$
near the left boundary at minus infinity.
This translates into an essentially constant magnetization profile with a fast transition 
layer near the left boundary. for $J>1$ and $h<0$ the picture is similar.

These arguments imply that in a first approximation ($L$ and $w$ infinite) the van der Waals curve of the chain-CW system is given by
the Maxwell construction of the single CW system. In order to get the finer structure around the Maxwell plateau we have to do a more careful finite size analysis.

\vskip 0.25cm
\noindent {\bf Asymptotics for $L>>w>>1$ large and $J\to 1_+$.}  
Now we set 
\begin{equation}
 t=\sqrt{J-1}x,\qquad \mu(x) = \mu(\frac{t}{\sqrt{J-1}}) \equiv \sqrt{J-1}\sigma(t)
\end{equation}
and look at the regime $J\to 1_+$. A straightforward calculation shows that the 
left hand side of equation \eqref{becomes} becomes
\begin{equation}\label{becomes2}
\frac{J(J-1)^{\frac{3}{2}}}{2}\biggl\{\int_{-\infty}^{+\infty} dx g(\vert x\vert ) x^2\biggr\}
\sigma^{\prime\prime}(t) + O((J-1)^{\frac{5}{2}}),
\end{equation}
and that the right hand side becomes
\begin{equation}\label{becomes3}
(J-1)^{\frac{3}{2}}(-\sigma(t) + \frac{1}{3}\sigma(t)^3) - h
 + O((J-1)^{\frac{5}{2}}).
\end{equation}
Lastly, we set $\tilde h =h (J-1)^{-\frac{3}{2}}$, and thus from \eqref{becomes}, 
\eqref{becomes2}, \eqref{becomes3}
\begin{equation}\label{motion}
\kappa\sigma^{\prime\prime}(t) = -\sigma(t) + \frac{1}{3}\sigma(t)^3 -\tilde h.
\end{equation}
Again, this is Newton's second law for a particle of mass $\kappa$ moving in the inverted potential
\begin{equation}\label{invert}
 V(\sigma) = \frac{1}{2}\sigma^2 - \frac{1}{12}\sigma^4 +\tilde h\sigma.
\end{equation}
The boundary conditions \eqref{critical} mean that the initial and final positions
of the particle for $t\to\pm\infty$ are the solutions of
\begin{equation}\label{of}
 \sigma_{\pm} - \frac{1}{3}\sigma_{\pm}^3 +\tilde h =0,
\end{equation}
corresponding to the local maxima of the potential.
Initial and final velocities are automatically fixed by 
the requirement that $\lim_{t\to\pm\infty}\sigma(t) = \sigma_{\pm}$.

Summarizing, in the limit  
\begin{equation}\label{limit}
 \lim_{J\to 1_{+}; h(J-1)^{-\frac{3}{2}}\,\text{fixed}}\,\lim_{w\to+\infty}\,\lim_{L\to+\infty}
\end{equation}
the magnetization profile is 
\begin{equation}\label{approxi}
 m_z\approx \sqrt{J-1} \sigma\bigl(\sqrt{J-1}\frac{z}{w}\bigr)
\end{equation}
where $\sigma(t)$ is a solution of \eqref{motion}.
\vskip 0.25cm
\noindent{\bf Kink states.} For $\tilde h=0$ (meaning $h=0$) \eqref{motion} has 
 the well known 
solutions 
\begin{equation}\label{tan}
 \sigma^{\rm kink}(t) = \sqrt{3}\tanh \biggl\{\frac{t-\tau}{\sqrt{2\kappa}}\biggr\}
\end{equation}
The center $\tau$ of the kink is a parameter
that we have to fix  from the normalization condition.
From \eqref{approxi} and \eqref{tan} we have
\begin{align}
\frac{1}{2L+1}\sum_{z=-L}^{+L} m_z & \approx 
 \frac{\sqrt{3(J-1)} }{2L+1}\sum_{z=-L}^{+L}
\tanh(L\frac{\sqrt{J-1}}{w\sqrt {2\kappa}}(\frac{z}{L}-\frac{w\tau}{L\sqrt{J-1}}))
\nonumber
\\
& \approx
\frac{\sqrt{3(J-1)}}{2}\int_{-\infty}^{+\infty} dx \,
{\rm sign}(\frac{\sqrt{J-1}}{w\sqrt {2\kappa}}(x-\frac{w\tau}{L\sqrt{J-1}}))
\nonumber
\\
& \approx
\frac{\sqrt{3}w\tau}{L}
\end{align}
Since this sum must be equal to $m$ we find
$\tau\approx\frac{mL}{\sqrt{3}w}$. The net result for the magnetization profile is
\begin{equation}\label{kinky}
m_z^{\rm kink}\approx \sqrt{3(J-1)}\tanh \biggl\{\frac{1}{w}\sqrt{\frac{J-1}{2\kappa}}
(z - \frac{mL}{\sqrt{3(J-1)}})\biggr\}
\end{equation}
\vskip 0.25cm
\noindent{\bf Homogeneous states.} When $\tilde h\neq 0$ the solution 
cannot be put in closed form. To lowest order in $\tilde h$ the solutions of \eqref{of} are 
$\sigma_{\pm}=\pm\sqrt 3 +\tilde h$. The initial velocity is (assuming the final velocity is zero)
to leading order,
\begin{equation}
 \sqrt{\frac{2}{\kappa}(V(\sigma_{+})-V(\sigma_{-}))}\approx 
2\frac{{3}^{1/4}}{\tilde h^{1/2}}
\end{equation} 
Thus, roughly speaking, the particle travels with constant 
velocity $2\frac{3^{1/4}}{\sqrt\kappa}\tilde h^{1/2}$ 
from position
$-\sqrt 3 +\tilde h$  
 during a finite time 
$O(\frac{3^{1/4}\sqrt \kappa}{h^{1/2}})$ and 
then stays exponentially close to the final position $\sqrt 3 +\frac{\tilde h}{2}$. 
The magnetization profile 
is 
\begin{equation} \label{Homy}
  m_z \approx 
\begin{cases}
-\sqrt{3(J-1)} +\frac{h}{2(J-1)} 
+ \frac{2}{\sqrt\kappa}(3(J-1))^{1/4} h^{1/2} (\frac{z+L}{w}) ,\\ \qquad\qquad\qquad\qquad\qquad -L\leq z\leq -L+
O(\frac{w\sqrt\kappa}{2(3(J-1))^{1/4}h^{1/2}})
\\
\\
\sqrt{3(J-1)} + \frac{h}{2(J-1)} ,\qquad\qquad z\geq -L+
O(\frac{w\sqrt\kappa}{2(3(J-1))^{1/4}h^{1/2}})
 \end{cases}
\end{equation}

\vskip 0.25cm
\noindent{\bf Comparison of free energies.}
In this paragraph we compute a naive approximation
for the free energy \eqref{cano-free-energy}. First consider the energy
difference $F_L^{\rm kink}-F_L^{\rm const}$ between kink $m_z^{\rm kink}$ and constant $m_z^{\rm const}=m$ states both with 
total magnetization $m_{-}< m < m_{+}$ on the Maxwell plateau.
We write this as 
\begin{equation}
 F_L^{\rm kink}-F_L^{\rm const} = (F_L(m_\pm)- F_L^{\rm const}) + 
(F_L^{\rm kink}-F_L(m_\pm))
\end{equation}
Because of \eqref{laplacian-sum} the first term is easily 
estimated as $(2L+1)(\Phi(m_{\pm})- \Phi(m)) + O(w)$ which is 
negative for $m$ on the Maxwell plateau. Since the magnetization density of
the kink state tends exponentially fast to $m_{\pm}$ for $z\to \pm\infty$ the second term is 
clearly $O(w)$ and therefore for $L$ large the kink states are stable\footnote{this argument breaks down 
for $\vert m-m_\pm \vert= O(L^{-\frac{1}{2}})$; this is discussed in sect. \ref{section-3}}.
But our interest here is in a precise calculation of this second term which displays 
an interesting oscillatory structure.
\begin{align}\label{bigsum}
F_L^{\rm kink}-F_L(m_\pm)= &-\frac{1}{2}\sum_{z,z^\prime=-L}^L 
D_{z,z^\prime} (m_z^{\rm kink} m_{z^\prime}^{\rm kink} - m_{\pm}^2)
\\
\nonumber
&
+ \sum_{z=-L}^{L} (\Phi(m_z^{\rm kink}) - \Phi(m_{\pm}))
\end{align}
Using \eqref{kinky} and \eqref{laplacian-sum} it is easy to 
see that, in the bulk, \eqref{bigsum} is a periodic
function of $m$ with period $\frac{\sqrt{3(J-1)}}{L}$, as long as the center of the kink is in the bulk. 
To compute it
we first extend the sums to infinity and use the Poisson summation formula
\begin{equation}
 \sum_{z\in \mathbb{N}} F(z) = \sum_{k\in \mathbb{N}} 
\int_{-\infty}^{+\infty} dz e^{2\pi i k z} F(z)
\end{equation}
for 
\begin{align}\label{F}
 F(z) = 
-\frac{1}{2}\sum_{z^\prime=-\infty}^{+\infty} 
D_{z,z^\prime} (m_z^{\rm kink} m_{z^\prime}^{\rm kink} - m_{\pm}^2) 
+ \Phi(m_z^{\rm kink}) - \Phi(\sqrt{3(J-1)})
\end{align}
A look at \eqref{kinky} shows that it has poles in the complex plane at 
$z_n= \frac{mL}{\sqrt {3(J-1)}} + i\pi (n+\frac{1}{2}) w\sqrt{\frac{2\kappa}{J-1}}$, 
$n\in \mathbb{N}$.  This suggests that the first term in \eqref{F} has the same pole structure.
The second term involving the potential is more subtle because its exact expression 
involves a logarithm which induces branch cuts.
However one can show, keeping the true expression for the potential, 
that
the branch cuts are outside of a strip 
$\vert {\Im(z)}\vert <\frac{\pi}{2}w\sqrt{\frac{2\kappa}{J-1}}$, and therefore
$F(z)$ is analytic in this strip. 
This is enough to deduce from  
standard Paley-Wiener theorems that for 
$w\sqrt{\frac{2\kappa}{J-1}}$ large
 $\vert F(k)\vert  =O(e^{-\vert k\vert w\sqrt{\frac{2\kappa}{J-1}}(\pi^2-\epsilon)})$.
In the appendix we perform a detailed analysis 
to show (for $J\to 1_+$, $w$ large and $k$ fixed)
\begin{equation}\label{assymp}
\int_{-\infty}^{+\infty} dz e^{2\pi i k z} F(z) \approx 4(J-1)\kappa w^2 \pi^2 k\bigl(1-k^2 \frac{\pi^2 w^2 \kappa}{J-1}\bigr)
\sinh^{-1}\biggl(k\pi^2 w \sqrt{\frac{2\kappa}{J-1}}\biggr)
\end{equation}
Retaining the dominant terms $k=0$ and $k=\pm 1$
in the Poisson summation formula we find for the free energy ($m_{-}<m<m_{+}$)
\begin{align}\label{main-result-1}
 F_L^{\rm kink}(m)\approx & (2L+1)\Phi(m_{\pm})  +
 4w(J-1)^{3/2} \sqrt{\frac{\kappa}{2}} \nonumber \\ &
 - 16(\pi w)^4\kappa^2 e^{-\pi^2 w\sqrt{\frac{2\kappa}{J-1}}
} 
\cos\biggl(2\pi  m \frac{L}{\sqrt {3(J-1)}}\biggr)
\end{align}
This result confirms the Maxwell construction, namely that the free energy per unit length converges 
to the convex envelope of $\Phi(m)$.
The finite size corrections display an interesting structure. The first correction $O((J-1)^{3/2})$ comes from the zero mode and 
represents the ''surface tension'' of the kink interface. The oscillatory term is a special feature 
of coupled mean field models. 
As explained in more details 
in section \ref{section-6}, general arguments 
show that 
for an anisotropic Kac limit such a convergence to 
the Maxwell plateau would not occur through 
oscillations but through a sequence of convex curves.
According to formula \eqref{kinky} 
$\frac{mL}{\sqrt{3(J-1)}}$ is the position of the kink, thus the profiles 
centered at integer positions correspond to 
minima of the periodic potential and are stable, while
those centered at half-integer positions correspond to 
maxima and are therefore unstable states.
The energy difference 
between a kink centered at an integer and one centered at a neighboring half-integer
is a Peierls-Nabarro barrier 
\begin{equation}
 32(\pi w)^4\kappa^2 e^{-\pi^2 w\sqrt{\frac{2\kappa}{J-1}}}.
\end{equation}
This is the energy needed to displace the kink along the chain. Such energy barriers are usually derived within 
effective soliton like equations for the motion of defects in crystals \cite{Nabarro}. Here the starting point was a microscopic
statistical mechanics model.
\vskip 0.25cm
\noindent{\bf Oscillations of the van der Waals curve.}
The van der Waals curve is easily obtained ($m_{-}<m<m_{+}$)
\begin{align}\label{main-result-2}
 h & = \frac{1}{2L+1} \frac{\partial F_L^{\rm kink}(m)}{\partial m}
= \frac{1}{2L+1} \frac{\partial}{\partial m}(F_L^{\rm kink} - F_L(m_\pm))
\\
\nonumber
&
\approx 
\frac{16\pi (\pi w)^4\kappa^2}{\sqrt{3(J-1)}} 
e^{-\pi^2 w\sqrt{\frac{2\kappa}{J-1}}} 
\sin\biggl(2\pi  m \frac{L}{\sqrt {3(J-1)}}\biggr)
\end{align}
At this point we note that the limit $L\to +\infty$ and $\frac{\partial}{\partial h}$ do not commute. 
This is so because on the Maxwell plateau we have a 
sequence of transitions\footnote{these can be thought as first order phase transitions with 
infinitesimal jump discontinuities} from one kink state to another.
In accordance with the numerical calculations, we find a curve that oscillates around the Maxwell plateau
$m\in [-\sqrt {3(J-1)}, +\sqrt {3(J-1)}]$ with a period $O(\frac{\sqrt{3(J-1)}}{L})$. 
The amplitude of these oscillations is exponentially small with respect to $w$ and thus much smaller
than the height $O((J-1)^{3/2})$ of the spinodal points (see \eqref{spin}). For example for the uniform 
coupling function we have $\kappa =1/6$ and the amplitude of the oscillations is 
$O(e^{-\pi^2 w\sqrt{\frac{1}{3(J-1)}}})$.
\vskip 0.25cm
\noindent{\bf Uniform interaction: $h=0$ and all $J$.} In case of a uniform 
interaction along the chain 
$g(\vert x\vert)=\frac{1}{2}$, $\vert x\vert \leq 1$ and $0$ 
otherwise, it turns out that equation \eqref{form} has 
the exact solution
\begin{equation}
\mu(x) = m_{\pm}\tanh Jm_{\pm} (x - x_0)
\end{equation}
for all $h=0$ and $J$.
This can be checked directly by inserting the 
function in \eqref{form} and 
seeing that it
reduces to the CW equation for $m_{\pm}$. Of course this 
solution is non trivial only for 
$J>1$. Relating the center $x_0$ to the total magnetization we get
the magnetization profile
\begin{equation}\label{specific}
m_z\approx m_{+}\tanh\biggl\{\frac{Jm_\pm}{w}(z- \frac{m}{m_{\pm}} L)\biggr\}
\end{equation}
Here $\approx$ means that $L>>w>>1$. One can check that the formula reduces 
to \eqref{kinky} when $J\to 1_+$. With this expression one can compute an 
exact formula for the exponent of the amplitude of 
oscillations of the van der Waals curve.
Indeed as argued after \eqref{F} this exponent is solely determined by the location of the poles 
of \eqref{specific} for $z\in \mathbb{C}$. Therefore we obtain for the 
case of the uniform interaction and all $J>1$, 
\begin{equation}\label{main-result-3}
 h = C(w,J) e^{-\frac{\pi^2 w}{Jm_\pm}} \sin\bigl(2\pi \frac{m}{m_{\pm}} L\bigr)
\end{equation}
where $C(J, w)$ is a prefactor that could in principle be computed by extending the calculation 
of the Appendix. Up to this prefactor, the Peierls-Nabarro barrier is $e^{-\frac{\pi^2 w}{Jm_+}}$ for all $J>1$.
\vskip 0.25cm
\noindent{\bf Remarks.} The main features of these 
oscillations, their period and exponentially small amplitude,  
are independent
of the details of
the exact model and its free energy. Only the prefactor will 
depend on such details. The period 
is equal to $\frac{m_{+}-m_{-}}{2L}$ where $m_+-m_{-}$ is the 
width of the Maxwell plateau.
The 
wiggles have an amplitude $e^{-2\pi \Delta}$ where $\Delta$ is 
the width of a strip in $\mathbb{C}$ where the 
kink profile is analytic (when the position variable $z$ is continued 
to $\mathbb{C}$). In general we have 
$\Delta= \alpha\frac{w\pi}{2Jm_{+}}$ were $\alpha= O(1)$. For the uniform window
$\alpha =1$ and in general when $J\to 1_+$ we have $\alpha\to \kappa^{-1/2}$. 
The point here is that the amplitude of the wiggles does not depend 
on the details of the free energy but only on the 
locations of the singularities $m_z^{\rm kink}$ in the complex plane.
If an explicit formula is not 
available for the kink profiles $\Delta$ can still be estimated by 
numerically computing
the discrete Fourier transform of the kink 
and identifying $\Delta$ with its rate of decay. 
This quantity will always be proportional 
to the scale factor $w$ in $m_z^{\rm kink}$.

\section{Random field coupled Curie-Weiss system}\label{section-5}

In this section we extend the problem to a chain of random field Curie-Weiss (RFCW) models.
The analysis being similar we only give the main steps for the simplest such extension. We will adopt the canonical formulation. Consider 
again a chain of single RFCW systems attached to
positions $z=-L,...,+L$. At each position we have a set of $N$ spins $s_{iz}$, $i=1,...,N$ interacting through the 
coupling \eqref{spin-spin-coupling} and subject to a random magnetic field $h_{iz}$. The r.v $h_{iz}$ are i.i.d with
zero mean $\mathbb{E}[h_{iz}]=0$ and for simplicity we assume that they take on a finite number of values
$\mathbb{P}(h_{iz}=H_\alpha) = p_\alpha$ where $\alpha$ runs over some finite index set. The Hamiltonian is 
\begin{equation}\label{RFCW}
H_{N,L}  = 
-\frac{1}{N}\sum_{\langle iz,jz^\prime\rangle} J_{z,z^\prime}
s_{iz} s_{jz^\prime}
-\sum_{i=1,z=-L}^{N,L} h_{iz}s_{iz}.
\end{equation}
The partition function involves
a sum over all configurations satisfying 
the constraint $\sum_{i,z=1,-L}^{N,L} s_{iz} = mN(2L+1)$. 

For each $z=-L,...,+L$ we introduce the variables 
\begin{equation}
  m_{z\alpha}=\frac{1}{\vert I_{z\alpha}\vert}\sum_{i\in I_{z\alpha}} s_{iz} 
\qquad \text{where}\qquad 
I_{z\alpha}=\{i\mid h_{iz} = H_\alpha\}.
\end{equation}
In terms of these the magnetization profile becomes
\begin{equation}
 m_z=\frac{1}{N}\sum_{i=1}^N s_{iz} = \sum_{\alpha}\frac{\vert I_{z\alpha}\vert}{N} m_{z\alpha},
\end{equation}
and the Hamiltonian (up to a constant term)
\begin{equation}\label{ham-RFCW}
 H_{N,L}=-\frac{N}{2}\sum_{z,z^\prime=-L}^L D_{z,z^\prime}
m_{z} m_{z^\prime}
-\frac{NJ}{2}\sum_{z=-L}^L m_z^2
-N\sum_{z=-L}^L\sum_\alpha \frac{\vert I_{z\alpha}\vert}{N} H_\alpha m_{z\alpha}.
\end{equation} 
When $N\to +\infty$ typical realizations 
of the random field satisfy $\frac{\vert I_{z\alpha}\vert}{N}\to p_\alpha$ and the Hamiltonian 
becomes deterministic.
The partition function can be expressed as an integral 
analogous to \eqref{compact-spins}, which yields for the canonical 
free energy
\begin{align}\label{rfcw-can}
F_{L}^{\rm r.f}(m) = &\min_{\sum_{z=-L}^L m_z = m(2L+1)}\biggl\{
 -\frac{1}{2}\sum_{z,z^\prime=-L}^L D_{z,z^\prime}
m_{z} m_{z^\prime}
\nonumber\\
&
-\sum_{z=-L}^L (\frac{J}{2}m_z^2 +\sum_\alpha p_\alpha 
(H_\alpha m_{z\alpha} + \mathcal{H}(m_{z\alpha}))\biggr\}
\end{align}
As usual the van der Waals curve is given by the 
relation $h=\frac{1}{2L+1}\partial F_{L}^{\rm r.f}(m)/\partial m$. 

To carry out the minimization in \eqref{rfcw-can} we apply the gradient operator
\begin{equation}
\bigl(\frac{d}{dm_{z\alpha}} = \frac{\partial}{\partial m_{z\alpha}} +
p_\alpha\frac{\partial}{\partial m_{z}}\,\,;\,\,\frac{\partial}{\partial\lambda}\bigr)
\end{equation}
to the associated Lagrangian and equate to zero.
This leads to the set of equations 
\begin{equation}\label{critic}
\begin{cases}
\sum_{z^\prime=-L}^L D_{z,z^\prime} m_{z^\prime}  
 =  -Jm_{z}-\lambda - H_\alpha + 
\frac{1}{2}\ln\bigl(\frac{1+m_{z\alpha}}{1-m_{z\alpha}}\bigr),\,\,\,\, {\rm all}\,\,\alpha
\\
m  =  \frac{1}{2L+1}\sum_{z=-L}^L\sum_\alpha m_{z\alpha}.
\end{cases}
\end{equation}
Similarly to \eqref{var}, if $(m_{z\alpha}^*, \lambda^*)$ is a solution, it can be 
shown that $\lambda^* = h$. Moreover 
multiplying
\eqref{critic} by $p_\alpha$ and summing over $z$ we obtain 
\begin{equation}
 h= -Jm +\frac{1}{2L+1}\sum_{z=-L}^L\sum_\alpha  
\frac{p_\alpha}{2}\ln\biggl(\frac{1+m_{z\alpha}^*}{1-m_{z\alpha}^*}\biggr) +O(\frac{w}{L})
\end{equation}
This generalizes \eqref{two}. 

Finally we note that from \eqref{critic} or working directly in a grand-canonical ensemble 
one can derive a generalized CW equation for the magnetization profile:
\begin{equation} \label{rcirt}
 m_z= \sum_\alpha p_\alpha
\tanh\bigl\{J m_z + h + H_\alpha + \sum_{z^\prime=-L}^L D_{z,z^\prime} m_{z^\prime}\bigr\}.
\end{equation}
The solutions $m_z^*(h)$ of this equation yield still another 
representation for the van der Waals curve $m= \frac{1}{2L+1}\sum_{z=-L}^L m_z^*(h)$.

In this case it is more difficult to analyze the continuum approximation and we rely on the numerical solutions of the next section. 
The picture which
emerges is essentially the same as in the deterministic model.

\section{Numerical solutions}\label{section-3}
We have carried out the numerical computations both for the equations in the canonical and grand-canonical formulations. 
These confirm the analytical predictions for the oscillations of the van der Waals curve. Near the end points of the Maxwell plateau 
the situation is not identical for the canonical and grand-canonical ensembles because boundary effects become important. For 
simplicity we start with the grand-canonical formulation.
\vskip 0.25cm
\noindent {\bf Grand-canonical equations.}
It is convenient to solve a slightly different system of equations than \eqref{cri} in order to eliminate boundary effects
(one may think of this as a modification of the model at the boundaries of the chain)
\begin{equation}\label{forced}
\begin{cases}
 m_z= \tanh\bigl\{Jm_z +h +\sum_{z^\prime=-L-w+1}^{+L+w-1} D_{z,z^\prime}  m_{z^\prime}\bigr\}, \quad -L \leq z \leq +L \\
\, m_{z}  = m_{+}(h), \quad L+1 \leq z \leq L+w-1 \\
\, m_{z}  = m_{-}(h), \quad -L-w+1 \leq z \leq -L-1 . \\
\end{cases}
\end{equation}
In other words, we force the profile to equal $m_-(h)$ at extra positions $-L-w+1$ to $-L-1$ and to $m_+(h)$ at extra 
positions  $L+1$ to $L+w-1$.
The van der Waals relation $h(m)$ is recovered from the solutions $m_z^*(h)$ of \eqref{forced} by using \eqref{total-mag}.
The first equation is equivalent to
\begin{equation}\label{69}
h= -(J + D_{zz})m_z + \tanh^{-1} m_z - \sum_{z^\prime=-L-w+1, z^\prime\neq z}^{L+w-1}
D_{zz^\prime} m_{z^\prime}
\end{equation}
Summing over $z$ and using \eqref{total-mag} we obtain
\begin{equation}\label{70}
 h= -(J + D_{zz})m + \frac{1}{2L+1}\sum_{z=-L}^{L}\biggl\{\tanh^{-1} m_z - \sum_{z^\prime=-L-w+1, z^\prime\neq z}^{L+w-1}
D_{zz^\prime} m_{z^\prime}\biggr\}
\end{equation}
Also, \eqref{69} is equivalent to
\begin{equation}
m_z(J+D_{z,z}-1)=
\tanh^{-1} m_z-m_z- \displaystyle \sum_{z'=-L-w+1, z'\neq z}^{L+w-1} D_{z,z^\prime}  
m_{z^\prime} -h
\end{equation}
The last two equations are the basis of:
\begin{algorithm}
{\small
\caption{Iterative solutions of \eqref{forced}}
\begin{algorithmic}[1]
\STATE Fix $m$. Initialize $m_z^{(0)}=m$ for $-L \leq z \leq L$ and $h^{(0)}=0$.
\STATE From $m_z^{(t)}$ compute: 
\begin{equation*}
h^{(t+1)}\leftarrow(J+D_{z,z})m+ \frac{1}{2L+1}\sum_{z=-L}^{L}\biggl\{\tanh^{-1} m_z^{(t)} - \sum_{z^\prime=-L-w+1, z^\prime\neq z}^{L+w-1}
D_{zz^\prime} m_{z^\prime}^{(t)}\biggr\}
\end{equation*}
\STATE For $-L \leq z \leq +L$, update $m_z^{(t+1)}$ as
\begin{equation*}
m_z^{(t+1)}\leftarrow
\frac{1}{J+D_{z,z}-1} 
\biggl\{\tanh^{-1} m_z^{(t)}-m_z^{(t)}- \displaystyle \sum_{z'=-L-w+1, z'\neq z}^{L+w-1} D_{z,z^\prime}  
m_{z^\prime}^{(t)} -h^{(t+1)}\biggr\}
\end{equation*}
and for a tunable value $\theta$ (for $\theta=0.9$ the iterations are ``smooth'')
\begin{equation*}
m_z^{(t+1)}\leftarrow\theta m_z^{(t)}+ (1-\theta)m_z^{(t+1)} 
\end{equation*}
\STATE For $-L-w+1 \leq z \leq -L-1$ let $m_z^{(t+1)}\leftarrow m_-(h^{(t+1)})$ and 
for $L+1 \leq z \leq L+w-1$ let $m_z^{(t+1)}\leftarrow m_+(h^{(t+1)})$.
\STATE Continue until $t=T$ such that the $\ell_1$ distance between the two consecutive profiles is less than some prescribed error $\delta$.
Output $h^{(T)}(m)$ and $m_z^{(T)}$.
\end{algorithmic}
}
\end{algorithm}

\begin{figure}[ht]
\begin{centering}
\input{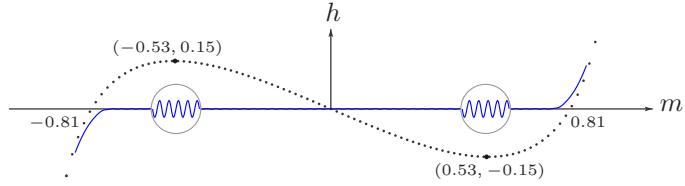}
\caption{ \small{Dotted line: van der Waals curve of single system for $J=1.4$. Continuous line: van der Waals isotherm for $J=1.4$, $L=25$, $w=1$ and 
$g(0) = \frac{1}{2}$, $g(\pm 1)= \frac{1}{4}$. Circles: $40$-fold vertical magnification. Throughout the plateau one has
$50$ wiggles corresponding to $50$ stable kink states.}
}
\label{VdW-L=25}
\end{centering}
\end{figure}

\begin{figure}[ht]
\begin{centering}
\input{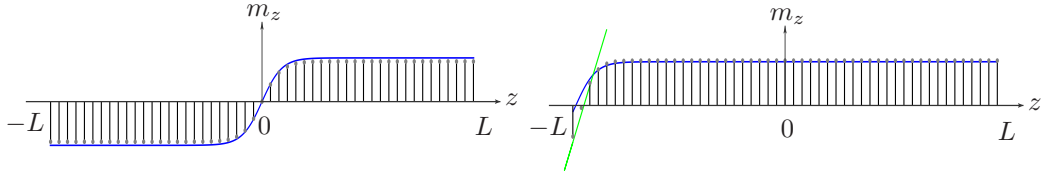}
\caption{ {\small Vertical bars are the numerical values and the continuous lines (blue and green) are given by equations \eqref{kinky}, \eqref{Homy}. Left: 
kink state centered at $m=0$ (so $h=0$) and $J=1.4$, $L=25$, $w=1$, 
$g(0) = \frac{1}{2}$, $g(\pm 1)= \frac{1}{4}$. Right: homogeneous solution for the same $J$, $L$, $w$, $g$ and $h(m)=0.017$.}
}
\label{const-J=1.1}
\end{centering}
\end{figure}

Figures \ref{VdW-L=25} and \ref{const-J=1.1} show the output of this procedure for $L=25$, $w=1$, 
$g(0) = \frac{1}{2}$, $g(\pm 1)= \frac{1}{4}$. We see from Figure~ \ref{const-J=1.1} that when $J=1.1$, already for $w=1$ the 
continuum approximation equ. \eqref{kinky} for the profile is good.  

Table~\ref{table:nonlin} compares the numerical 
amplitude of the oscillations $N_w$ for the van der Waals curve with 
the analytical formula \eqref{main-result-2}
\begin{equation}
 \underbrace{\frac{16\pi (\pi w)^4 \kappa^2}{\sqrt{3(J-1)}}}_{C_w}\underbrace{\exp\biggl(-\pi^2 w\sqrt\frac{2\kappa}{J-1}\biggr)}_{E_w}.
\end{equation}
We take $J=1.05$, and the triangular window $g(\vert x\vert)=\frac{2w}{1+3w}(1-\frac{\vert x\vert}{2})$. In order to get a stable result
for $w=3$ we have to go to lengths $L=250$. We see that the agreement is quite good for 
the exponent while the prefactor seems to be off by a constant factor $O(1)$. 

\begin{table}
\caption{{\small Amplitude of wiggles: $J=1.05$ and triangular window.}}
\vskip 0.15cm
\centering
\begin{tabular}{c c c c c c c}
\hline\hline 
$w$ & $N_w$ & $E_w$ & $C_w$ & $\frac{\log N_w}{ \log C_wE_w}$ & $\frac{\log \frac{N_w}{C_w} }{ \log E_w}$ & $\frac{\log \frac{N_w}{E_w} }{ \log C_w}$  \\ [0.5ex]
\hline
$1$ & $2.5 \times 10^{-12}$ & $ 2.8 \times 10^{-14}$ & $7.9\times 10^{2}$ & $1.09$ &  $1.07$ & $0.67$\\
$2$ & $3.4 \times 10^{-22}$ & $ 9.3 \times 10^{-25}$ & $7.8 \times 10^{3}$ & $1.07$ & $1.06$ & $0.66$ \\ 
$3$ & $6.7 \times 10^{-32}$ & $ 5.1 \times 10^{-35}$ & $3.2 \times 10^{4}$ & $1.05$ & $1.04$ & $0.69$  \\
$4$ & $3.2 \times 10^{-41}$ & $ 3.3 \times 10^{-45}$ & $9.2 \times 10^{4}$ & $1.02$ & $1.02$ & $0.80$  \\
\hline
\end{tabular}
\label{table:nonlin}
\end{table} 

For larger values of $J$ and uniform window $g(\vert x\vert) = \frac{w}{2w+1}$ we can use formula \eqref{main-result-3} 
to compare the numerical amplitude $N_w$ with 
$E_w= e^{-\frac{\pi^2 w}{Jm_\pm}}$. Table~\ref{table:nonlin2} shows the results for $J=1.4$ and $L=80$.

\begin{table}
\caption{{\small Amplitude of wiggles: $J=1.4$ and uniform window.}}
\vskip 0.15cm
\centering
\begin{tabular}{c c c c}
\hline\hline 
$w$ & $N_w$ & $E_w$ & $\frac{\log N_w}{ \log E_w}$   \\ [0.5ex]
\hline
$1$ & $2.2 \times 10^{-5}$ & $ 1.7 \times 10^{-4}$ & $1.24$\\
$2$ & $3.5 \times 10^{-9}$ & $ 3.0 \times 10^{-8}$ & $1.12$  \\ 
$3$ & $5.9 \times 10^{-13}$ & $ 5.2 \times 10^{-12}$ & $1.08$  \\
$4$ & $1.0 \times 10^{-16}$ &  $ 9.0 \times 10^{-16}$ & $1.06$ \\
\hline
\end{tabular}
\label{table:nonlin2}
\end{table}

\vskip 2cm
\noindent{\bf Canonical equations.}
Let us now discuss the numerical solutions of \eqref{crit}. Here the boundary conditions are not forced at the outset
and adjust themselves to non-trivial values when $m$ is on the plateau. It turns out that for some values of $m$ the  
output of iterations is greatly affected by the choice of the initial 
profile. Thus in 
order to find the correct global minimum of the canonical
free energy a suitable initial condition  must be chosen. A natural choice is to choose 
the solution of \eqref{forced} as the initial point. The numerical procedure is as 
follows:
\begin{algorithm}
{\small
\caption{Iterative solutions of \eqref{crit}}
\begin{algorithmic}[1]
\STATE Fix $m$. Initialize $m_z^{(0)}$ and $h^{(0)}$ to a solution of \eqref{forced} given by algorithm 1. 
\STATE  From $m_z^{(t)}$ compute: 
\begin{equation*}
h^{(t+1)}\leftarrow (J+D_{z,z})m-\frac{1}{2L+1}\displaystyle \biggl\{\sum_{z=-L}^L \tanh^{-1} m_z^{(t)} + \displaystyle \sum_{z=-L}^L 
\displaystyle \sum_{z'=-L, z' \neq z}^{L} D_{z,z^\prime}  m_{z^\prime}^{(t)}\biggr\} 
\end{equation*}
\STATE For $-L \leq z \leq +L$, first update $m_z^{(t+1)}$ as:
\begin{equation*}
m_z^{(t+1)}\leftarrow 
\frac{1}{J+D_{z,z}-1}\biggl\{\tanh^{-1} m_z^{(t)} -m_z^{(t)}- \sum_{z'=-L, z'\neq z}^{L} D_{z,z^\prime}  m_{z^\prime}^{(t)} -h^{(t+1)}\biggr\}  
\end{equation*}
and for a tunable value $\theta$ (say $\theta=0.9$),
$$
m_z^{(t+1)}\leftarrow \theta m_z^{(t)}+ (1-\theta)m_z^{(t+1)}
$$
\STATE Continue until $t=T$ such that the $\ell_1$ distance between the two consecutive profiles is less than a prescribed error $\delta$. Output
$h^{(T)}$ and $m_z^{(T)}$.
\end{algorithmic}
}
\end{algorithm}

\begin{figure}[h!]
\begin{centering}
\input{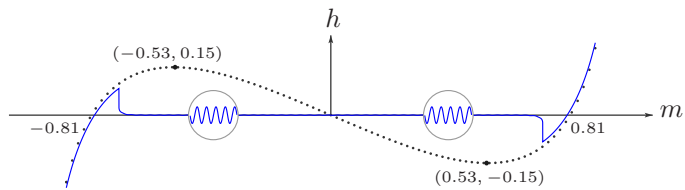}
\caption{{\small Dotted line: isotherm of single system for $J=1.4$. Continuous line: isotherm of coupled model with $L=25$, $w=1$,
$g(0)= \frac{1}{2}$, $g(\pm)=\frac{1}{2}$. Vertical magnification factor in the circle is $40$.  
For $\vert m-m_{\pm}\vert = O(L^{-\frac{1}{2}})$ there is a boundary effect explained in main text.}
}
\label{cano-VdW-L=25}
\end{centering}
\end{figure}

\begin{figure}[h!]
\begin{centering}
\input{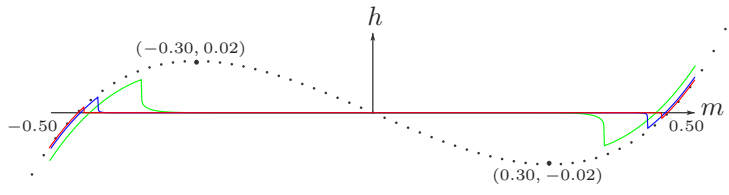}
\caption{{\small Behavior of the boundary effect for $J=1.1$ (same $w$ and $g$ as above) and $L=25$, $100$, $400$.}
}
\label{cano-differentL}
\end{centering}
\end{figure}

\begin{figure}[h!]
\begin{centering}
\input{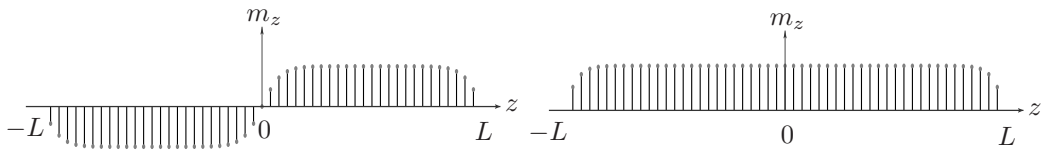}
\caption{ {\small Magnetization profiles for $J=1.1$, $L=25$ (same $w$ and $g$ as above). 
Left: kink centered at $m=0$. Right: homogeneous solution for $h(m)=0.017$.}
}
\label{cano-const-J=1.1}
\end{centering}
\end{figure}

Figure~\ref{cano-VdW-L=25} shows the van der Waals curve  
for $J=1.4$ with $L=25$, $w=1$ and $g(0)= \frac{1}{2}$, $g(\pm 1)=\frac{1}{2}$. Apart from the usual oscillations on the Maxwell plateau we observe that near
the extremities (close to $m_\pm$) the curve follows the metastable branch of the single system. This can easily be explained 
from equ. \eqref{main-result-1}. Indeed, the energy difference between a kink and constant state ($m_z=m$) is
\begin{equation}
 (2L+1)\Phi(m_\pm) + 4w (J-1)^{3/2}\sqrt{\frac{\kappa}{2}} - (2L+1)\Phi(m)
\end{equation}
where we drop the exponentially small oscillatory contribution. When $\vert m-m_\pm\vert$ is very small this difference becomes positive
because of the surface tension contribution of the kink, and the constant state is the stable state. It is easily seen that this happens for
$(m- m_\pm)^2 < \frac{2w}{2L+1}\sqrt{\frac{\kappa}{2}}(J-1)^{3/2}$
As seen in
fig.~\ref{cano-differentL} this boundary effect vanishes as $L$ grows large. Finally fig. \ref{cano-const-J=1.1} displays magnetization
 profiles: in the bulk they are identical to the grand-canonical ones, while near the boundaries the magnetization is 
reduced since the effective 
ferromagnetic interaction is smaller.

\noindent{\bf Random field coupled CW model.}
The numerical procedures and solutions are similar to the deterministic model so for simplicity we only consider a grand-canonical 
formulation with forced boundary conditions for extra vertices at the two extremities of the chain: 
\begin{equation}\label{forced-random}
\begin{cases}
 m_z= \sum_\alpha p_\alpha
\tanh\bigl\{J m_z + h + H_\alpha + \sum_{z^\prime=-L-w+1}^{L+w-1} D_{z,z^\prime} m_{z^\prime}\bigr\}, -L \leq z \leq +L \\
\, m_{z}  = m_{+}(h), \quad L+1 \leq z \leq L+w-1 \\
\, m_{z}  = m_{-}(h), \quad -L-w+1 \leq z \leq -L-1 ,\\
\end{cases}
\end{equation}
where $m_+(h)$ and $m_-(h)$ are the two stable solutions of a RFCW equation
$m= \sum_\alpha p_\alpha\tanh\bigl\{J m + h + H_\alpha \bigr\}$.
We have implemented the following iterative procedure:
\begin{algorithm}
{\small
\caption{Iterative solution of \eqref{forced-random}}
\begin{algorithmic}[1]
\STATE Fix $m$. Initialize $m_z^{(0)}=m$ for $-L \leq z \leq L$ and $h^{(0)}=0$.
\STATE Update $h^{(t+1)}\leftarrow X$ where $X$ is the (unique) solution of:
\begin{equation*}
 \frac{1}{2L+1}{\sum_{z=-L}^{L} \sum_\alpha p_\alpha \tanh\bigl\{J m_z^{(t)} + X + H_\alpha + \sum_{z^\prime=-L-w+1}^{L+w-1} D_{z,z^\prime} 
m_{z^\prime}^{(t)}\bigr\}} =   m.
\end{equation*}
\STATE For $-L \leq z \leq +L$, update $m_z^{(t+1)}$ as:
\begin{multline*}
m_z^{(t+1)}\leftarrow 
\frac{1}{J+D_{z,z}-1}   \sum_\alpha p_\alpha \biggl\{(J+D_{z,z})m_z^{(t)}\\ -
 \tanh\bigl\{J m_z^{(t)} + h^{(t+1)} + H_\alpha + 
\sum_{z^\prime=-L-w+1}^{L+w-1} D_{z,z^\prime} 
m_{z^\prime}^{(t)}\bigr\} \biggr\}.
\end{multline*}
and then: 
$m_z^{(t+1)}\leftarrow\theta m_z^{(t)}+ (1-\theta)m_z^{(t+1)}$ (for a tunable $\theta$).
\STATE For $-L-w+1 \leq z \leq -L-1$ let $m_z^{(t+1)}\leftarrow m_-(h^{(t+1)})$ and for $L+1 \leq z \leq L+w-1$ 
let $m_z^{(t+1)}\leftarrow m_+(h^{(t+1)})$.
\STATE Continue until $t=T$ s.t the $\ell_1$ distance between two consecutive profiles is less than a prescribed error $\delta$.
Output $h^{(T)}$ and $m_z^{(T)}$.
\end{algorithmic}
}
\end{algorithm}

\begin{figure}[ht]
\begin{centering}
\input{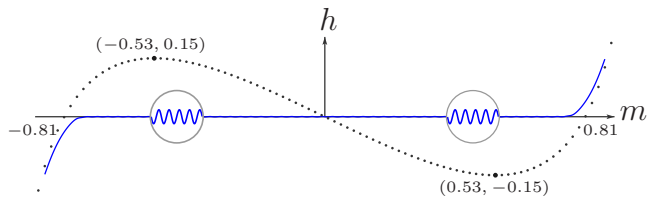}
\caption{ {\small Dotted line: RFCW isotherm for $J=1.4$. Continuous line: isotherm of coupled model 
for $J=1.4$, $L=25$, $w=1$ and $g(0) = \frac{1}{2}$, $g(\pm 1)= \frac{1}{4}$. In the circle: vertical magnification factor is $160$.}
}
\label{rand-VdW-J=1.4}
\end{centering}
\end{figure}

In Figure~\ref{rand-VdW-J=1.4} we illustrate the output $h^{(T)}$ of these iterations for case where 
the random field takes two values $H_{\pm} = \pm 0.1$ with probabilities $p_\pm = \frac{1}{2}$.  

\section{Anisotropic Kac limits}\label{section-6}

In the chain CW models the convergence of the van der Waals free energy and isotherm to the Maxwell construction 
is through a sequence of oscillatory curves (see equs \eqref{main-result-1} and \eqref{main-result-2}). This phenomenon is intimately related to 
the fact that the reference model that gets coupled is mean field. This may be a complete graph model or a 
sparse graph model;  it has to be, in some sense, an infinite dimensional system (see section \ref{section-7}). 
The oscillatory behavior is not found in anisotropic Kac limits of finite dimensional models. 
This can be understood very easily from the following arguments.

Suppose one considers a 2-dimensional Ising model on a rectangular lattice of dimensions $(2L+1)\times N$. Let $(r_0, r_1)$ be the vertices 
with $r_0\in -L,...,L$ the longitudinal component and $r_1\in 1,...,N$ the transverse component. Consider the spin-spin 
interaction
\begin{equation}
 \frac{J}{w_{\perp}w_{//}}\chi(w_{//} (r_0-r_0^\prime), w_{\perp} (r_1-r_1^\prime)) s_{(r_0r_1)} s_{(r_0^\prime r_1^\prime)}
\end{equation}
where $\chi(r_0,r_1) = 1$ for $\vert r_0\vert \leq 1$, $\vert r_1\vert\leq 1$ and $0$ otherwise. We consider the anisotropic case
$w_\perp>>w_{//}$ were a spin couples to many more spins along the transverse direction than in the longitudinal one. 
Note that for $w_{\perp}= N$ one recovers a coupled chain CW model and the transverse direction becomes effectively infinite dimensional. Our interest here is for $w_\perp>>w_{//}$ both of $O(1)$ with 
respect to $N$ and $L$. One can consider two types of anisotropic Kac limits, namely 
\begin{equation}
\lim_{w_{//}}\lim_{w_{\perp}}\lim_{L}\lim_{N}\qquad {\rm and} \qquad
\lim_{w_{//}}\lim_{L}\lim_{w_{\perp}}\lim_{N}
\end{equation}
where it is understood that all parameters tend to $+\infty$. In both cases the first 
limit $\lim_N$ can be thought of, as the thermodynamic limit of a quasi-1-dimensional strip of fixed width $2L+1$. 
Let $f_{N,L}(m)$ be the canonical 
free energy per spin of this quasi-1-dimensional model. Since the interaction is finite range we know by general principles 
that $\lim_{N\to+\infty} f_{N,L}$
is a convex function of $m$ for any fixed $L$, $w_\perp$, $w_{//}$. Since the limits of convex functions are convex
(when they exist) both remaining limiting processes will happen through a sequence of convex functions with no oscillatory behavior. 
Likewise the van der Waals curve
of the anisotropic model will in both cases converge to the Maxwell plateau but only through a sequence of 
increasing functions with no oscillations.

The same discussion applies to a $d+1$ dimensional model on a box of size $(2L+1)\times N^d$. But one can also consider 
the following variation: instead of $w_\perp\to+\infty$ take $d\to+\infty$. When $d\to+\infty$ is done after $N\to\infty$ there will 
be no oscillatory behavior. On te other hand when $d\to+\infty$ is done before $N\to\infty$, then the transverse direction effectively 
becomes a Bethe lattice and we get a coupled chain of 
mean field systems which will display an oscillatory behavior.

\section{Conclusion}\label{section-7}

We introduced the present model as a ''toy model`` to understand the threshold saturation phenomenon that is at the root of the 
excellent performance of  
recent constructions in the field of error correcting
codes for noisy channel communication. As mentioned briefly in the introduction the same phenomenon occurs in a wide variety of 
spatially coupled systems such as constraint satisfaction problems, compressed sensing and multi-user communication systems. 

Let us say a few words about the chains of constraint satisfaction problems \cite{Hassani-Macris-Urbanke-I}, \cite{Hassani-Macris-Urbanke-II}.
These include K-satisfiability, XOR-SAT, Q-coloring on Erdoes-Renyi graphs. The mean field 
solution of these systems (for the uncoupled system) 
is obtained by the cavity or replica methods \cite{Mezard-Parisi}. This solution is also closely related to message 
passing algorithms such as belief and/or survey propagation which predict
the existence of a region in the phase diagram with exponentially many metastable states between two thresholds: 
the ''survey propagation`` threshold and the SAT-UNSAT
phase transition threshold. By analyzing the cavity equations, for the coupled models with appropriate boundary conditions, we 
discover that, as the range of the Kac interaction grows, the survey propagation threshold saturates towards the SAT-UNSAT threshold.
This fact may have important algorithmic consequences that remain to be investigated.

What is the generic picture that emerges ? All systems considered above are coupled chains of 
individual infinite dimensional systems or mean field systems. Indeed the individual systems are defined on sparse graphs or complete graphs, which are both, in some sense, 
infinite dimensional objects. Besides, their exact (or conjecturally exact) solutions are given by mean field equations (Curie-Weiss equation, cavity/replica equations ect...).
These equations (for the individual system) have two stable fixed point solutions which describe the order parameter of the 
equilibrium states for the individual system. When boundary conditions are fixed such that the order parameter takes the two equilibrium values 
at the ends of the chain,  the spatially coupled system has a series of new equilibrium states corresponding to kink profiles. Since the kink interface is well localized
its free energy is
close to a convex combination of the two free energies corresponding to the boundary conditions. Because of the discrete nature of the chain there are tiny
free energy barriers corresponding to unstable positions for the kinks in-between two positions on the chain. This is the origin of the wiggles, 
both in the free energy functional (of CW or Landau or Bethe type) and in the van der Waals like curves.
A somewhat related discussion can also be found in \cite{tanaka}.
Let us note that this picture is exactly confirmed by an analysis of the weight enumerator and growth rate 
for the number of codewords of given relative weight of spatially coupled LDPC codes \cite{Hassani-Macris-Urbanke-Mori}.

It seems that the threshold saturation phenomenon should be quite generic among all $\infty+1$ dimensional systems which support kink-like equilibrium states.
There are many open questions that are worth investigating. For example, connections to coupled map systems, discrete soliton equations and the 
stability of their solutions, would allow to better understand when the phenomenon occurs or does not. Also the algorithmic 
implications of the phenomenon of threshold saturation is a largely open issue.

\section{Appendix}

We give the main steps leading to formulas \eqref{assymp} and \eqref{main-result-1}. First we notice that 
\begin{equation}\label{integral}
 \widehat F(k)\equiv \int_{-\infty}^{+\infty} dz\, e^{2\pi i kz} F(z) = e^{2\pi ik\frac{mL}{\sqrt{3(J-1)}}} \int_{-\infty}^{+\infty} dz e^{2\pi i kz} G(z)
\end{equation}
with 
\begin{align}\label{G}
 G(z) =  -\frac{1}{2}\sum_{z^\prime=-\infty}^{+\infty} 
D_{z,z^\prime} (m_z^{0} m_{z^\prime}^{0} - m_{\pm}^2) 
+ \Phi(m_z^{0}) - \Phi(\sqrt{3(J-1)})
\end{align}
and $m_z^{0}$ is a kink centered at the origin,
\begin{equation}
 m_z^0 = \sqrt{3(J-1)} \tanh\biggl\{\frac{1}{w}\sqrt{\frac{J-1}{2\kappa}} z\biggr\}.
\end{equation}
Now we evaluate the sum over $z^\prime$ in 
the first term of \eqref{G}. Setting $z^\prime = w x^\prime$ we have for $w$ very large, 
\begin{align}
 \sum_{z^\prime=-\infty}^{+\infty} 
 D_{z,z^\prime} m_{z^\prime}^{0} 
 & = \frac{J\sqrt{3(J-1)}}{w}\sum_{z^\prime=-\infty}^{+\infty} 
(g(\vert \frac{z}{w} - \frac{z^\prime}{w}\vert) - w \delta_{\frac{z}{w}, \frac{z^\prime}{w}})
\tanh\biggl\{\sqrt{\frac{J-1}{2\kappa}} \frac{z^\prime}{w}\biggr\}
\nonumber \\ &
\approx J\sqrt{3(J-1)}\int_{-\infty}^{+\infty} dx^\prime (g(\vert x^\prime\vert)-\delta(x^\prime))
\tanh\biggl\{\sqrt{\frac{J-1}{2\kappa}} (x^\prime + \frac{z}{w})\biggr\}
\nonumber \\ &
\approx J\sqrt{3(J-1)}\kappa
w^2\frac{d^2}{dz^2}\tanh\biggl\{\frac{1}{w}\sqrt{\frac{J-1}{2\kappa}} z \biggr\}
\end{align}
Therefore
\begin{equation}
 \sum_{z^\prime=-\infty}^{+\infty} 
 D_{z,z^\prime} m_{z^\prime}^{0} = -\sqrt{3}J(J-1)^{3/2} 
\bigl(1- \tanh^2\bigl\{\frac{1}{w}\sqrt{\frac{J-1}{2\kappa}} z \bigr\}\bigr)
\tanh\bigl\{\frac{1}{w}\sqrt{\frac{J-1}{2\kappa}} z \bigr\}
\end{equation}
In a similar way one shows that the $-m_{\pm}^2$ term does not contribute, and one finds
\begin{align}
 G(z) & \approx \frac{3}{2}J(J-1)^{2} 
\bigl(1- \tanh^2\bigl\{\frac{1}{w}\sqrt{\frac{J-1}{2\kappa}} z \bigr\}\bigr)
\tanh^2\bigl\{\frac{1}{w}\sqrt{\frac{J-1}{2\kappa}} z \bigr\}
\nonumber \\ &
+ \Phi
\bigl(\sqrt{3(J-1)} \tanh\bigl\{\frac{1}{w}\sqrt{\frac{J-1}{2\kappa}} z\bigr\}\bigr) - 
\Phi\bigl(\sqrt{3(J-1)}\bigr)
\end{align}
Replacing in \eqref{integral} we get after a scaling,
\begin{equation}\label{fourier-integral}
 \widehat F(k) = w\sqrt{\frac{2\kappa}{J-1}}e^{2\pi ik\frac{mL}{\sqrt{3(J-1)}}} 
\int_{-\infty}^{+\infty} dz e^{2\pi i kw\sqrt{\frac{2\kappa}{J-1}} z} 
\widetilde{G}(z)
\end{equation}
where 
\begin{align}\label{gtilde}
 \widetilde G(z)  \approx & \frac{3}{2}J(J-1)^{2} 
\bigl(1- \tanh^2 z\bigr)\tanh^2 z 
\nonumber \\ &
+ \Phi
\bigl(\sqrt{3(J-1)} \tanh z\bigr) - 
\Phi\bigl(\sqrt{3(J-1)}\bigr)
\end{align}
As a function of $z\in \mathbb{C}$, $\widetilde G(z)$ is analytic in the open strip
$\vert \Im(z)\vert<\frac{\pi}{2}$. Indeed $\tanh z$ has poles at 
$z_n=(n+\frac{1}{2})i\pi$,
$n\in\mathbb{Z}$ and $\Phi$ has branch 
cuts for $\sqrt{3(J-1)} \tanh z\in ]-\infty, -1]\cup[1, +\infty[$, or equivalently 
on the intervals
\begin{equation}
 z\in \cup_{n\in\mathbb{Z}} \biggl[z_n, z_n 
- \frac{1}{2} {\rm sign}(n) \ln\biggl\vert
\frac{1+\sqrt{3(J-1)}}{1-\sqrt{3(J-1)}}\biggr\vert\biggr].
\end{equation}
It is easy to see that the integrand in \eqref{fourier-integral} tends 
to zero exponentially fast, as $R\to +\infty$, 
for $z=\pm R + iu{\rm sign}(k)$, 
$\vert u\vert \leq \frac{\pi}{2}-\delta$ (any $0<\delta< 1$). Therefore we can shift the 
integration over $\mathbb{R}$ to the line $z= t+ i(\frac{\pi}{2}-\delta){\rm sign}(k)$,
$t\in \mathbb{R}$, which yields,
\begin{align}
 \widehat F(k) = & w\sqrt{\frac{2\kappa}{J-1}}e^{2\pi ik\frac{mL}{\sqrt{3(J-1)}}} 
e^{-\vert k\vert w \sqrt{\frac{2\kappa}{J-1}}\pi(\pi - 2\delta)}
\\ \nonumber &
\times
\int_{-\infty}^{+\infty} dt e^{2\pi i tw\sqrt{\frac{2\kappa}{J-1}}} 
\widetilde{G}(t+ i(\frac{\pi}{2}-\delta){\rm sign}(k))
\end{align}
From expression \eqref{gtilde} it is possible to show the 
estimate (for $\vert J-1\vert <<1$ and $0<\delta<<1$ and $C$ a numerical constant) 
\begin{equation}
 \vert \widetilde{G}(t+ i{\rm sign}k(\frac{\pi}{2}-\delta))\vert \leq 
C (J-1)^2 e^{-2\vert t\vert}\delta^{-4}.
\end{equation}
Since $\delta$ can be taken as small as we wish, this allows to conclude that 
\begin{equation}
 \widehat F(k) = C_{\delta,J,w}(k)\delta^{-4}(J-1)^{3/2} 
w\sqrt{2\kappa}e^{2\pi ik\frac{mL}{\sqrt{3(J-1)}}} 
e^{-\vert k\vert w \sqrt{\frac{2\kappa}{J-1}}\pi(\pi - 2\delta)}
\end{equation}
where $C_{\delta,J}(k)< C$ for all $k$. 
This result implies that the Van der Waals curve 
has oscillations, around the Maxwell plateau, of 
period $\frac{\sqrt{3(J-1)}}{L}$ and amplitude 
$e^{- w \sqrt{\frac{2\kappa}{J-1}}\pi^2}$.

By replacing the first terms of the expansion of $\Phi$ 
when $J\to 1_+$. we can obtain a completely explicit approximation 
for $\widehat F(k)$.  Thanks to the 
exact formula
\begin{equation}\label{GR}
 \int_{-\infty}^{+\infty} dz e^{ikz} (1-\tanh^4 z) 
= \frac{\pi}{6} \frac{k(8-k^2)}{\sinh \frac{k\pi}{2}}
\end{equation}
and using
$\Phi(m)\approx 
-\frac{J-1}{2} m^2 + \frac{1}{12} m^4$ we get 
\begin{equation}
\widetilde G(z)\approx\frac{3}{4}(J-1)^2
\bigl(1-\tanh^4  z \bigr),
\end{equation}
we find asymptotically for $w$ large, $J\to 1_+$ and any fixed $k$
\begin{equation}
\widehat F(k)\approx 4(J-1)\kappa w^2 \pi^2 k\bigl(1-k^2 \frac{\pi^2 w^2 \kappa}{J-1}\bigr)
\sinh^{-1}\biggl(k\pi^2 w \sqrt{\frac{2\kappa}{J-1}}\biggr).
\end{equation}
This is formula \eqref{assymp} of the main text. For the zero mode $k=0$ we get
\begin{equation}
 \widehat F(0)\approx 4(J-1)^{3/2} w \sqrt{\frac{\kappa}{2}}
\end{equation}
and for the other ones $k\in \mathbb{Z}^*$
\begin{equation}
 \widehat F(k) \approx -8 (\pi w)^4\kappa^2 \vert k\vert^3  
e^{-\vert k\vert \pi^2 w\sqrt{\frac{2\kappa}{J-1}}} 
e^{2\pi i k \frac{mL}{\sqrt {3(J-1)}}} 
\end{equation}

Finally, for the reader's convenience, we point out that to check \eqref{GR} one can use $\frac{1}{6}(\tanh z)^{\prime\prime\prime} + \frac{8}{6} (\tanh z)^\prime
= 1-\tanh^4 z$ and $\int_{-\infty}^{+\infty} dz e^{ikz} \tanh z = i\pi (\sinh \frac{\pi k}{2})^{-1}$ \cite{Gradstein-Ryzhik}.

\vskip 0.5cm

\noindent {\bf Acknowledgments.} N. Macris thanks C. E. Pfister and J. L. Lebowitz for discussions. We also 
thank J. H.H Perk and J. L. Lebowitz for giving pointers to the literature. The work 
of H. Hassani has been supported by
a grant of the Swiss National Science Foundation no 200021-121903.

\end{document}